\documentclass[aps,prd,twocolumn,amsmath,superscriptaddress,amssymb,showpacs,floatfix,nofootinbib,longbibliography]{revtex4-1}
\pdfoutput=1
\usepackage{graphicx}
\usepackage{bm}
\usepackage{times}
\usepackage{slashed}
\usepackage{color}
\usepackage{aas_macros}
\usepackage{slashed}
\usepackage{lipsum}
\usepackage{subfigure}
\usepackage{multirow}
\usepackage{amsmath}
\usepackage{array} 
\usepackage{varwidth} 
\usepackage{epsfig,psfrag,amssymb}
\usepackage{dcolumn}
\usepackage{hyperref}
\usepackage{ulem}
\usepackage{mathrsfs,amsfonts,hepunits}
\usepackage[utf8]{inputenc}
\usepackage{braket}

\newcommand{\be}{\begin{equation}}
\newcommand{\ee}{\end{equation}}
\newcommand{\bea}{\begin{eqnarray}}
\newcommand{\eea}{\end{eqnarray}}

\begin{document}

\title{Where are the Cascades from Blazar Jets? An Emerging Tension in the $\gamma$-ray sky}
\author{Carlos Blanco}
\email{carlos.blanco@fysik.su.se, ORCID: orcid.org/0000-0001-8971-834X}
\affiliation{Stockholm University and The Oskar Klein Centre for Cosmoparticle Physics,  Alba Nova, 10691 Stockholm, Sweden}
\affiliation{Princeton University, Department of Physics, Princeton, NJ 08544}
\author{Oindrila Ghosh}
\email{oindrila.ghosh@fysik.su.se, ORCID: orcid.org/0000-0003-2226-0025}
\affiliation{Stockholm University and The Oskar Klein Centre for Cosmoparticle Physics,  Alba Nova, 10691 Stockholm, Sweden}
\author{Sunniva Jacobsen}
\email{sunniva.jacobsen@fysik.su.se, ORCID: orcid.org/0000-0001-8732-577X}
\affiliation{Stockholm University and The Oskar Klein Centre for Cosmoparticle Physics,  Alba Nova, 10691 Stockholm, Sweden}
\author{Tim Linden}
\email{linden@fysik.su.se, ORCID: orcid.org/0000-0001-9888-0971}
\affiliation{Stockholm University and The Oskar Klein Centre for Cosmoparticle Physics,  Alba Nova, 10691 Stockholm, Sweden}

\begin{abstract}
\noindent Blazars are among the most powerful accelerators and are expected to produce a bright TeV $\gamma$-ray flux. However, TeV $\gamma$-rays are attenuated by interactions with intergalactic radiation before reaching Earth. These interactions produce cascades that transfer TeV power into the GeV band, powering both extended halos around bright sources and a large contribution to the isotropic $\gamma$-ray background (IGRB). Using conservative blazar models and recent IGRB measurements, we rule out scenarios where blazars effectively transfer their multi-TeV power into GeV $\gamma$-rays. Three possible solutions include: (1) strong spectral cuts on bright blazars, which are increasingly in tension with local blazar data, (2) collective plasma effects that can prevent the development of blazar cascades, the effectiveness of which is debated, (3) an increase in the $\gamma$-ray opacity from axion-like particles. 
\end{abstract}

\maketitle

Blazars, active galactic nuclei whose jets are oriented towards Earth, are powerful multiwavelength accelerators whose emission spans from MHz radio waves~\cite{2015Ap&SS.357...75M} to TeV \mbox{$\gamma$-rays}~\cite{Senturk:2013pa} and neutrinos~\cite{IceCube:2018cha}. However, the TeV $\gamma$-ray emission from distant blazars is attenuated by pair production on the extragalactic background light (EBL), composed of infrared and optical photons, as well as the CMB. Simultaneously, these pairs will inverse-Compton scatter (ICS) the same EBL and CMB photons to produce GeV $\gamma$-rays. This process proceeds continuously and converts a large fraction of the TeV $\gamma$-ray power into a detectable GeV $\gamma$-ray flux.

In the case that the intergalactic magnetic fields (IGMF) are weak, this signal could be observed as a diffuse ``pair halo" surrounding bright blazars. These searches are powerful, because a pair halo detection would serve as a smoking gun for TeV $\gamma$-ray production and attenuation, particularly if a time-delay between variations in the blazar point source and halo intensity were observed~\cite{Fermi-LAT:2018jdy}. However, no definitive signal has been detected in Fermi-LAT data~\cite{2011A&A...526A..90N, Chen:2014rsa, Chen:2018wck}, which could be explained if the IGMF is strong enough to sufficiently isotropize the signal such that no halos are observed~\cite{Neronov:2010gir,Tiede:2017aql}.

The blazar cascade contribution to the isotropic $\gamma$-ray background (IGRB), on the other hand, is guaranteed and stems from two components: (1) pair halos around blazars that are not bright enough to be detected, (2) pair halos surrounding detected blazars that are too disperse to be identified~\cite{Broderick:2013hir,Broderick:2011av,Murase:2012df,Fermi-LAT:2012nqz,Cholis:2013ena,Ajello:2011zi}. Importantly, if no pair halos are detected, the blazar contribution to the IGRB is maximally bright. Unfortunately, diffuse blazar contributions to the IGRB cannot be used to definitively detect $\gamma$-ray attenuation, due to the multitude of other sources that can contribute, most notably misaligned active galactic nuclei (mAGN) and star-forming galaxies (SFGs)~\cite{Fields:2010bw, Fermi-LAT:2012nqz, Hooper:2016gjy}. However, IGRB observations can be used to constrain the blazar contribution at GeV energies. Very recently, we have shown in Ref.~\cite{Blanco:2021icw} that mAGN and SFGs, along with a sub-dominant contribution from undetected blazar point sources, are expected to produce $\sim$100\% of the IGRB, leaving little room for a diffuse blazar cascade component.

\begin{figure}
    \centering
    \includegraphics[width=0.45\textwidth]{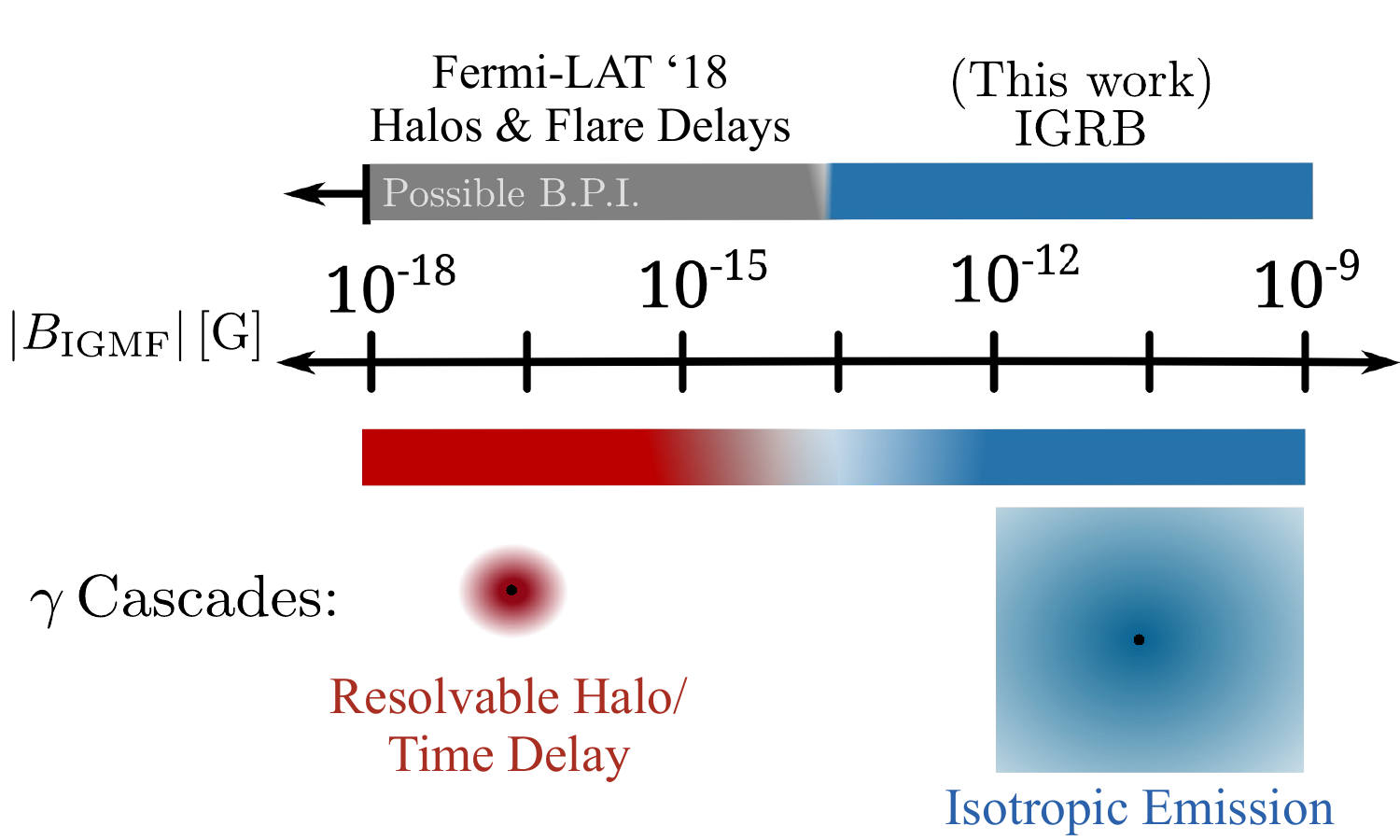}
    \vspace{-0.3cm}
    \caption{A diagram of the cascaded $\gamma$-ray flux from blazars as a function of the IGMF strength at a coherence length taken to be roughly $\lambda  \sim 1$~kpc. When the IGMF is weaker than $\sim10^{-14}$~G, the cascades are detectable as degree-scale halos and as a time delay in flares~\cite{Fermi-LAT:2018jdy}. When the IGMF is stronger, the cascades are essentially isotropic and contribute to the IGRB. For particularly weak fields, beam-pair instabilities could cool pairs before they generate cascades.}
    \vspace{-0.2cm}
    \label{fig:blazarSketch}
\end{figure}

The argument of this paper is as follows. Recent observations in several separate domains have crystallized each portion of the above story, proving that:

\begin{enumerate}
    \vspace{-0.2cm}
    \item Local blazars produce a bright TeV flux, and do not generally have sharp spectral cutoffs at $\mathcal{O}$(TeV) energies.
    \vspace{-0.2cm}
    
    \item The IGMF is too weak to cool $e^{+}e^{-}$ pairs and prevent them from producing bright $\gamma$-ray emission via ICS.
    
    \vspace{-0.2cm}
    \item Models of mAGN, SFG, and sub-threshold blazar point source dominate the IGRB, limiting the contribution of diffuse blazar cascade emission to fall below $\sim$10\%, which is in strong tension with blazar models.
    \vspace{-0.2cm}
\end{enumerate}

The only known Standard Model solution stems from the possibility that beam-plasma instabilities cool pairs before ICS can occur ~\cite{Broderick:2011av,Broderick:2018nqf,Alawashra:2022all,Chang:2013hia}. The efficiency of these instabilities depends sensitively on the injected pair spectrum, and the onset of saturation for unstable modes. Most critically, their efficiency is separated into several regimes based on the IGMF strength. For a benchmark correlation length of $\lambda_{B} \sim 1~\mathrm{kpc}$, instabilities remove significant energy for \mbox{$B < 10^{-18}~\mathrm{G}$}, but are negligible for \mbox{$ B \gtrsim 10^{-14}~\mathrm{G}$.} The hierarchy between ICS- and instability-driven losses is indeterminant (but sliding) in the intermediate range. Thus, our analysis provides strong evidence that the IGMF is feeble.

\noindent {\it Intrinsic Gamma-Ray Emission from Blazars -- }
Both theoretical and observational considerations indicate that blazars are bright TeV sources. Most notably, recent studies have identified a population of ``extreme TeV blazars", which produce $\gamma$-ray emission extending to at least 10~TeV~\cite{Biteau:2020prb}. Notably, these systems have high-energy synchrotron peaks that exceed 4~eV. This correlation is well-explained within the standard synchrotron self-Compton model~\cite{1998MNRAS.301..451G}, where the TeV $\gamma$-ray emission stems from the ICS of the same pairs that produce the synchrotron flux. This correlation allows for observations of the synchrotron peak to predict the intrinsic properties of the TeV flux. Observations indicate that nearly 25\% of the blazar population produces an extreme synchrotron peak~\cite{Chang:2019vfd}, with many systems detected at energies up to $\sim$5--10~keV. This implies that there is a significant population of blazars with intrinsic $\gamma$-ray emission up $\sim$1~PeV, which could dominate the IGRB, despite often being dim GeV point sources~\cite{2010MNRAS.401.1570T}.

Recently, a combination of H.E.S.S., VERITAS, and HAWC observations are beginning to verify the existence of nearby objects with detectable TeV $\gamma$-ray emission despite the presence of significant $\gamma$-ray attenuation. 
Most importantly, nearby blazars such as Mrk 421 and Mrk 501, which have been studied extensively, have been detected at up to 9 and 12 TeV respectively by several telescopes such as HAWC, VERITAS, H.E.S.S. and MAGIC~\cite{Acciari:2011kj, VERITAS:2010vjk, HAWC:2021obx}, and even higher energies during blazar flares \cite{Aharonian:1999vy, HESS:2005ndo}. Several other nearby sources have been detected above 6 TeV, including 1ES 1218+304, 1ES 1959+650 and 1ES 2344+514~\cite{Feng:2021sdq}. Note that the sources 1ES 1218+304 and 1ES 1959+650 have been observed at up to $\sim 10$~TeV and $\sim 15$~TeV, respectively. In addition, many sources that have not been detected at energies exceeding TeV energies show signs of hard spectra indicating that they emit $\gamma$-rays at the $\mathcal{O}(10)$ TeV range~\cite{MAGIC:2019ipe}. In particular, the majority of nearby sources ($z < 0.5$) that have been detected in the TeV range are defined as high-frequency synchrotron peaked, indicating that they emit $\gamma$-rays well into the $\mathcal{O}(10)$~TeV range. \\

\noindent {\it Gamma-Ray Attenuation and Blazar Cascades -- } These direct blazar point-source observations significantly under-represent the total power of TeV blazars, because the attenuation of TeV $\gamma$-rays by the EBL hides the $\gamma$-ray signal from all but the closest systems. This makes the majority of TeV blazars invisible to even sensitive searches for their point source emission at TeV energies.

Fortunately, the signature of the bright TeV $\gamma$-rays is imprinted on the GeV $\gamma$-ray spectrum from even distant blazars. Electromagnetic cascades are produced through the interaction of TeV-scale $\gamma$-rays with the EBL and CMB photons. The electron-positron pairs produced in these cascades will subsequently produce GeV $\gamma$-rays through the ICS off these same radiation fields. We note that ICS losses dominate over synchrotron losses so long as the magnetic field strength strength is below $\sim$1$\mu$G, which holds for all existing IGMF models~\cite{Planck:2015zrl, Amaral:2021mly}. Because the pairs are charged, they are first deflected by the IGMF, diffusing their signal compared to the point-like emission observed for $\gamma$-rays produced directly by the blazar.

This process can be broken down into two regimes based on the IGMF strength. When the IGMF is very weak \mbox{($\lesssim$ 10$^{-14}$~G)}, pairs are only slightly deflected from their original trajectories. This produces a diffuse ``pair halo" at GeV energies around the position of bright blazars, which could be detected in extended source searches. In addition to morphological searches, such pair halos could also be detected through searches for time-delays in the flux of highly-variable blazars. However, studies using Fermi-LAT data have not detected any blazar pair halos through either morphological or time-domain searches, constraining field strengths smaller than about $B\lesssim10^{-15}$ G for correlation lengths of about $\lambda_{B} \sim 1~\mathrm{kpc}$~\cite{Fermi-LAT:2018jdy} .

For stronger IGMFs \mbox{($B \gtrsim$10$^{-14}$~G)} , the pairs are deflected too significantly to be associated with any blazar -- and no pair halos will be observed. In this case, the entirety of the cascaded emission will then be observed as an contribution to the IGRB, which will be maximally bright. It is important to note that, for both small and intermediate IGMF strengths, the total GeV $\gamma$-ray flux is the same and only the division between the intensity of detectable pair halo and the IGRB contribution varies. Given that no pair halos have been observed, the expected TeV blazar contribution to the IGRB is relatively independent of the IGMF strength. 

Thus, the null observation of pair halos implies that the TeV $\gamma$-ray power from blazars is efficiently converted into isotropic GeV $\gamma$-rays that contribute to the IGRB. \\

\noindent {\it Beam-Plasma Instabilities -- } 
The only known exception to the above scenario stems from the possibility that interactions between the cold pair beam and the surrounding intergalactic medium (IGM) plasma, known as pair-beam plasma instabilities, effectively cool the TeV pairs before they can generate GeV emission through ICS. 
While the non-observation of pair halos may be attributed to efficient beam-plasma instabilities, this scenario requires magnetic fields weaker than a critical IGMF strength of $B \sim 10^{-18}-10^{-14}~\mathrm{G}$ for correlation lengths of about $\lambda_{B} \sim 1~\mathrm{kpc}$~\cite{Alawashra:2022all}. \\

\noindent {\it Modeling and Source Selection -- } 
We calculate the cascaded blazar contribution to the IGRB by modeling and adding the emission from blazars identified at GeV energies. Our blazar selection is based on Ref.~\cite{Yang:2022shw}, which calculated the spectral energy distribution of 2709 Fermi blazars from the 4FGL catalog~\cite{Fermi-LAT:2018jdy}. Out of these, we only include the 1788 sources that have defined redshifts. Furthermore, we cross check sources from Ref.~\cite{Yang:2022shw} with the 4FGL catalog, and find that 86 sources cannot be cross-identified, leaving us with 1702 simulated sources.

For each source, we extract the observed Fermi spectrum and the best-fit parameters at the pivot energy reported in 4FGL. Starting from these best-fit parameters, we refit the observed Fermi spectra of the blazars to one of three spectral models, a simple power law, a log parabola, or a power law with a super-exponential cutoff, as provided in Ref.~\cite{Fermi-LAT:2018jdy}. For the fitting, we take the Fermi best-fit parameters at the pivot energy as a seed, in order to ensure that we get a fit that can reproduce the observed flux.

We then correct for $\gamma$-ray attenuation to find the intrinsic $\gamma$-ray spectrum at the blazar source. 
The EBL-corrected data points are found by dividing by the average photon survival probability, $\langle P_{\gamma\gamma} \rangle$, in each energy bin, which is calculated as:

\begin{equation}
     \langle P_{\gamma\gamma}\rangle = \frac{\int_{\Delta E} \exp[-\tau_{\gamma \gamma}(z, E)]\phi_{\rm obs}(E)}{\int_{\Delta E} \phi_{\rm obs}(E)} \ ,
\end{equation}
where $\tau_{\gamma \gamma}\sim \int{d\epsilon \sigma_{\gamma\gamma} dn/d\epsilon } $ is the optical depth, $\sigma_{\gamma\gamma}$ is the pair-production cross section, $dn/d\epsilon (z)$ is the number density of the target background photons (EBL+CMB), and $\phi_{\rm obs}(E)$ is the observed flux, which we describe using either a power law, a log parabola, or power law with a super-exponential cutoff.

\begin{figure}[t]
    \centering
    \includegraphics[width=0.49\textwidth]{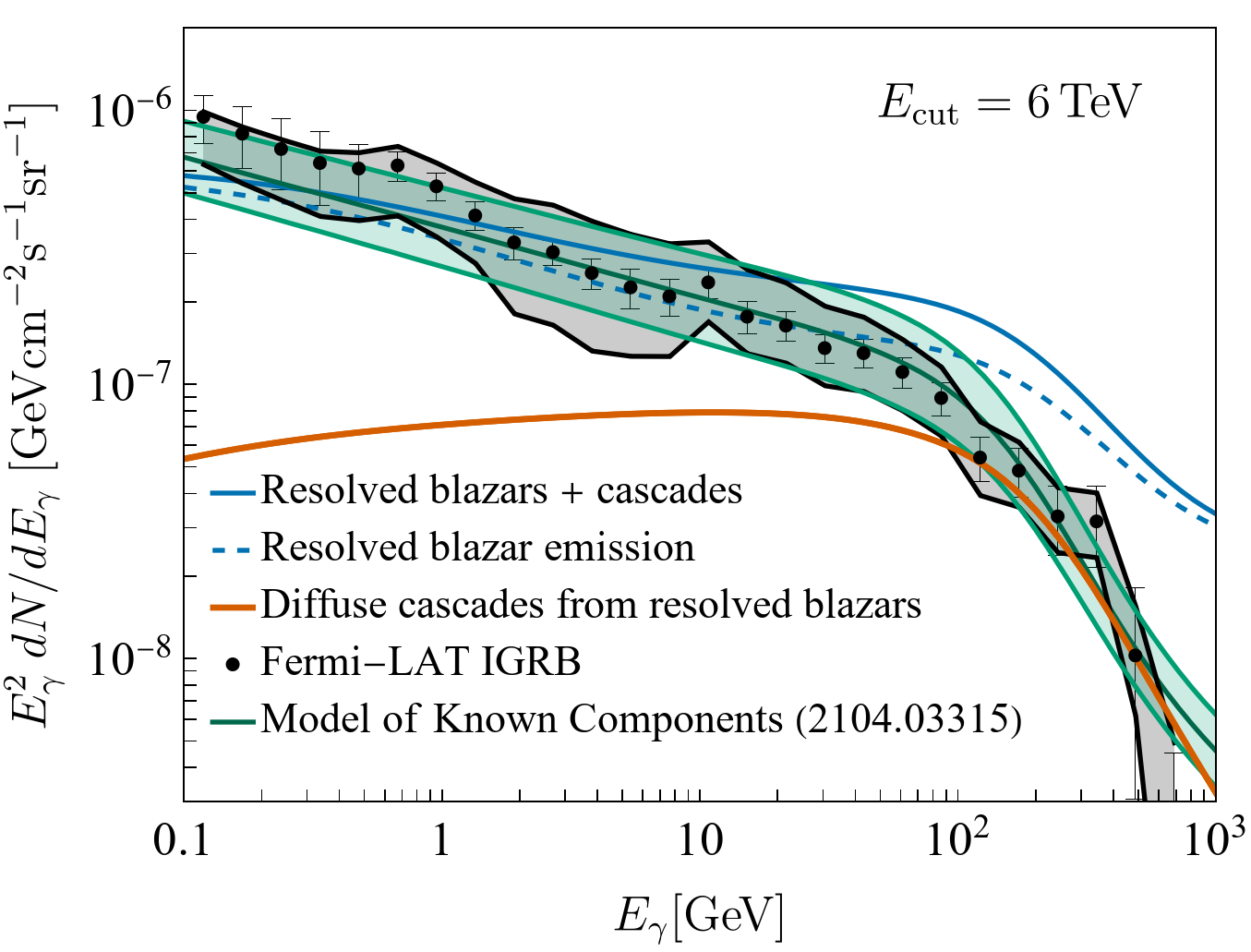}
    \includegraphics[width=0.49\textwidth]{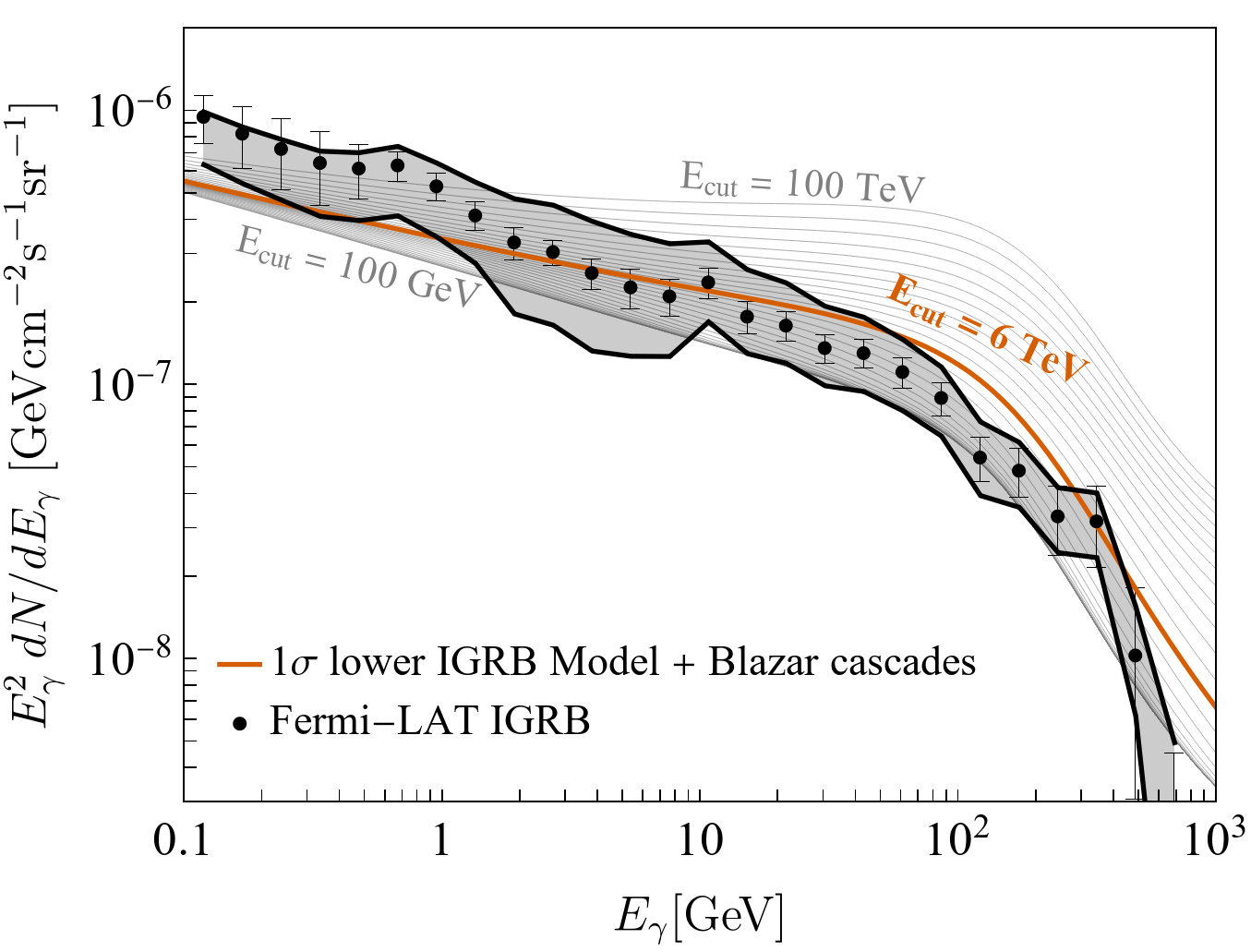}
    \caption{The top frame shows the all-sky flux from resolved blazars compared to the measured IGRB~\cite{Fermi-LAT:2014ryh} (black) and the model of known contributions~\cite{Blanco:2021icw} (green). The blue (dashed) line is the sum of the line-of-sight blazar fluxes while the blue (solid) line is the sum of the line-of-sight blazar fluxes plus their corresponding diffuse cascades. In (red) we show the diffuse cascades from the resolved blazars  for $E_{\rm cut}=6\, \rm TeV$. The bottom frame shows the expected diffuse $\gamma$-rays from the known contributions ---taken at the 1$\sigma$ lower limit--- plus the diffuse cascades from resolved blazars. We show thin lines for $E_{\rm cut}$ values between $100\, \rm GeV$ and $100\, \rm TeV$. We highlight the estimate for the maximum $E_{\rm cut}$ in red.}
    \label{fig:maxContribution}
\end{figure}

The intrinsic (de-absorbed) spectra are then re-fit to a spectral model to extrapolate them above 100 GeV. We note that sources with best-fit spectra with positive energy curvature are instead fit to power laws in order to assume the most conservative emission models above Fermi energies. We note that the overwhelming majority of the brightest sources (192/200) were best fit by downward-curving spectra. Of all sources, $\sim 27\%$ are fit by log parabolas or power laws with a super-exponential cutoff. \\

\noindent {\it EBL Models and Cascade Development -- } 
Here, we adopt the EBL model of Ref.~\cite{Dominguez:2010bv}. We use the publicly available code, \texttt{$\gamma$-cascade}~\cite{Blanco:2018bbf} to compute the continuous development of the electromagnetic cascade cycle from each source in our catalog. We also account for the effects of cosmological redshift on the $\gamma$-ray propagation. Since we consider only IGMF strong enough to isotropize the produced pair's trajectory, the  cascades are essentially isotropic. Finally, in order to infer the intrinsic spectrum of the blazars, we reverse the effect of line-of-sight attenuation due to pair production. \\

\noindent {\it Known components of the IGRB - }
Over the last two decades, observations of astrophysical $\gamma$-rays from the Fermi-LAT~\cite{Fermi-LAT:2014ryh}, have produced precise measurements of the IGRB. Specifically, these are photons originating from outside of the Milky Way that are not attributed to known or resolved point sources. The IGRB measurement is generated by first constructing the extragalactic $\gamma$-ray background (EGRB), which includes all extragalactic $\gamma$-rays regardless of origin,  and then masking known sources. 

Previous studies have concluded that the IGRB can almost entirely be accounted for by $\gamma$-rays generated by mAGN and SF activity in unresolved galaxies~\cite{Hooper:2016gjy,Linden:2016fdd,Blanco:2021icw}. There is also a contribution coming the point source emission of unresolved blazars, which produces a significant portion of the diffuse $\gamma$-ray flux below $\sim$10 GeV and potentially above 100 GeV. However, it should be noted that previous studies of unresovled blazars have not fully taken into account the cascaded component of blazar jet emission~\cite{Cholis:2013ena,DiMauro:2013zfa} which introduces a significant uncertainty in the unresolved blazar contribution. While Ref.~\cite{DiMauro:2013zfa} computes a flux from cascaded photons, they account only for the CMB and do not include the IR and optical components of the EBL. This approximation systematically underpredicts the cascaded component of the blazar contribution because the evolution of electromagnetic cascades for photons with energies below around 100~TeV is dominated by interactions with the EBL~\cite{Skrzypek:2022hpy}. In this study, we remain agnostic about the exact makeup of the IGRB due to the fact that uncertainties in the composition of the IGRB are insignificant compared to the potential magnitude of diffuse cascades from resolved blazars. We therefore adopt the IGRB contribution model derived in Ref.~\cite{Blanco:2021icw}, and note that due to the fact that the blazar contribution is expected to be underestimated, our results are conservative. \\

\begin{figure}[t]
    \centering
    \includegraphics[width=0.49\textwidth]{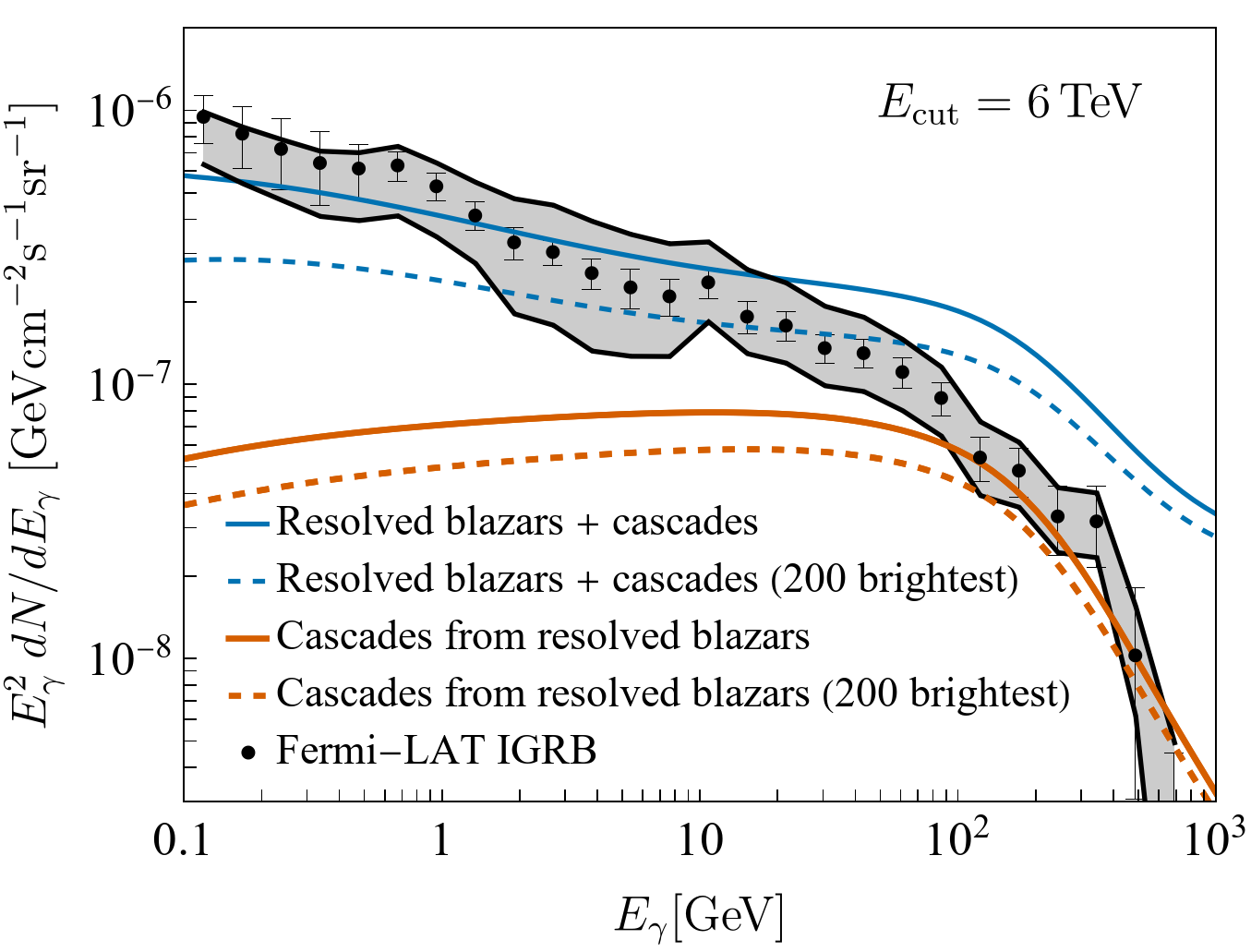}
    \caption{The relative contribution of the 200 brightest blazars is shown in the (dashed) lines. The (blue) lines show the sum of the line-of-sight blazar fluxes plus their corresponding diffuse cascades. The (red) lines show the diffuse cascades from the resolved blazars for $E_{\rm cut}=6\, \rm TeV$. The 200 brightest blazars contribute about half of the diffuse GeV cascades.}
    \label{fig:brightest}
\end{figure}

\noindent{\it Results --- } Figure~\ref{fig:maxContribution} shows the main result of this study, namely that if blazars inject significant amounts of energy above $\sim$5 TeV, they would overproduce the observed IGRB. We introduce an additional parameter corresponding to an exponential cutoff ($E_{\rm cut}$) into the intrinsic spectra of all blazars. We then compute the line-of-sight, and cascaded $\gamma$-ray flux as a function of $E_{\rm cut}$. We find that if $E_{\rm cut}\gtrsim 6\,\rm{TeV}$, the isotropic cascades exceed the measured IGRB when added to the model of known components.

In the bottom frame of Figure~\ref{fig:maxContribution}, we show the total predicted isotropic $\gamma$-ray flux for a range of blazar cutoff energies. We use the $1\sigma$ lower range for the model of known IGRB components (from Ref.~\cite{Blanco:2021icw}) in order to produce conservative constraints on our maximum $E_{\rm cut}$ estimate. Note that while the exact value of the maximum $E_{\rm cut}$ may be uncertain at the level of $\mathcal{O}(1)$, the expected diffuse $\gamma$-ray flux from blazar cascades quickly overproduces the entire IGRB regardless of the model of known IGRB components. In order to demonstrate the robustness of our analysis, we show the contribution from the 200 brightest Fermi-LAT blazars in Figure~\ref{fig:brightest}, finding that these sources produce roughly half of the total emission. Therefore, in the event of a luminosity-dependent cutoff that prevents either the brightest or dimmest blazars from  generating cascades, our main results would remain approximately the same.  \\

\noindent{\it Effect of Beam-Plasma Instabilities --- } 
Our results indicate that the diffuse cascaded emission should overproduce the Fermi IGRB at GeV energies, unless the blazar population includes an relatively low $\gamma$-ray spectral cutoff that is inconsistent with theoretical models of blazar emission and in tension with current TeV observations. The most promising alternative method within Standard Model physics for dimming the GeV signal is through energy losses owing to beam-plasma instabilities as the charged pairs propagate through the IGM plasma. The efficiency of such energy losses is debated, since the cooling rates depend on the growth rate of the unstable modes, ranging from the reactive instability growth characterized by a short cooling time, to the slower kinetic modes which fall behind by several orders of magnitude \citep{brejzman1974powerful,Miniati:2012ge}. The parameters of concern that determine this efficiency through scaling relations are the beam-plasma density contrast, beam Lorentz factor, and beam temperature. In particular, the energetic width of the pair distribution function, known as beam temperature, at injection determines whether the unstable modes have a reactive or kinetic growth, which in turn affects the aforementioned scaling relations.

The exponential growth of unstable modes can be sketched using linear stability analysis, and the estimates obtained are valid until Landau damping and other nonlinear effects become important, leading to saturation. Significant inhomogeneities in the IGM \citep{Perry:2021rgv} or instability-driven electrostatic fluctuations within the beam \citep{Sironi:2013qfa} have the potential to induce energetic broadening of the pair beam, leading to its stabilization \citep{Ghosh:2023tevlab}, preventing the instabilities from removing energy from the beam and cooling the pairs. Furthermore, weak tangled IGMF $B_{IGM} > 10^{-18}~\mathrm{G}$ with pc-scale correlation lengths can lead to an angular broadening of the pair beam, effectively quenching the growth of unstable electrostatic Langmuir modes \citep{Alawashra:2022all} capable of suppressing the cascades. Based on this estimate, very feeble IGMF are required in order for these instabilities to effectively cool the pair beam.

The above scenario applies to slow kinetic growth rates for a generic Gaussian pair beam distribution function at injection \citep{Miniati:2012ge}, where the instability-associated energy loss time \mbox{$\tau_{\mathrm{loss}} \sim 1 / (2 \delta_{\mathrm{growth}} \omega_{p})$} is comparable to the inverse Compton timescale $\tau_{\mathrm{ICS}}$. However, we note that for a relativistic Maxwellian, e.g., a Maxwell-J\"uttner beam distribution function \citep{bret2010exact,Broderick:2011av,Chang:2016gji} or the less realistic case of a monochromatic beam \citep{fainberg1972nonlinear}, the corresponding cooling time $\tau_{\mathrm{loss}}$ for TeV pair beams is significantly shortened. In these cases, field strengths of $B_{\mathrm{IGM}} \gtrsim  10^{-14}~\mathrm{G}$ and $B_{\mathrm{IGM}} \gtrsim  10^{-11}~\mathrm{G}$ are required, respectively. Therefore, the efficiency of the beam-plasma instability is sensitive to the chosen pair beam distribution function at injection, which can be constructed from a realistic initial pair spectrum at production \citep{Schlickeiser:2012hn,Vafin:2018kox}, introducing a corresponding sliding scale in the range of $B_{\mathrm{IGM}} \sim 10^{-18}-10^{-14}~\mathrm{G}$, as the critical IGMF strength necessary to stabilize the pair beam, for a correlation length of $\lambda_{B} \sim 1~\mathrm{kpc}$. 

The above discussion is based on linear theory estimates and confined to the consideration of electrostatic modes. In addition, beam-Weibel instabilities \citep{weibel1959spontaneously} can lead to current filamentation and magnetization of plasma. However, energetic broadening in the direction parallel to beam propagation can erase the thermal anisotropies necessary to trigger such unstable electromagnetic modes. We refrain from forecasting the evolution of the beam due to nonlinear effects through which the growth of the unstable modes saturate, as simulations investigating regimes beyond the linear growth phase are not viable for realistic parameters at the moment.

\noindent{\it Conclusions --- } In this letter, we have shown that conservative models of blazar spectra, TeV $\gamma$-ray cascade physics, and IGMF magnitude predict that the observed blazar population should produce a bright, isotropic GeV $\gamma$-ray signal which is ruled out by existing Fermi-LAT data. We introduce this as an emerging ``tension", which requires some modification of our present astrophysical models. The most convincing standard model explanation is that beam-plasma instabilities efficiently cool the pairs before they produce GeV $\gamma$-ray emission through ICS. This has significant implications for models of plasma physics in the IGM, and for the strength of the IGRB.

 The implications of this study go beyond blazar and IGRB modeling. If blazars do have strong intrinsic spectral cutoffs in their TeV emission, their correlated TeV and PeV neutrino emission would be strongly suppressed, strengthening claims that blazars do not significantly contribute to the IceCube neutrino flux~\cite{IceCube:2016qvd, Smith:2020oac, Hooper:2018wyk}. This, in turn, strengthens the case for hidden neutrino sources~\cite{Capanema:2020oet,Murase:2015xka}. Secondly, any additional blazar contribution would further saturate the astrophysical contributions to the IGRB and significantly enhance current IGRB constraints on both annihilating~\cite{Fermi-LAT:2015qzw, DiMauro:2015tfa}, and more importantly, decaying dark matter models~\cite{Blanco:2018esa}(see also constraints on MACHOs~\cite{Graff:1999gv}).

Finally, an intriguing solution to the tension introduced in this paper is the mixing between axion-like particles, or ALPs, and photons in external magnetic fields, described by the Lagrangian: $\mathcal{L}_{a\gamma} = g_{a\gamma}a\boldsymbol{E}\cdot\boldsymbol{B}$, where $a$ is the ALP and $g_{a\gamma}$ is the ALP-photon mixing strength. This effect may lead to a decreased $\gamma$-ray flux from blazars, if a significant fraction of the $\gamma$-rays are converted into ALPs. ALP-photon mixing can happen in two scenarios. One is that ALP-photon mixing in the blazar jets leads to an increased TeV transparency, which in turn leads to an energy leakage away from the cascades. 
Another is that ALP-photon mixing leads to an increased GeV opacity. These effects will be investigated further in a future paper~\cite{upcomingaxionpaper}. \color{black}

\label{sec:nulltest}

\section*{Acknowledgements}
We would like to thank Avery Broderick, Christopher Eckner, Katie Freese, Dieter Horns and Kohta Murase for comments which improved the quality of this manuscript. We would especially like to thank Mattia Di Mauro and Alexander Korochkin for making us aware of a numerical error which affected the blazar contribution in the original draft of this paper. The work of C.B.~was supported in part by NASA through the NASA Hubble Fellowship Program grant HST-HF2-51451.001-A awarded by the Space Telescope Science Institute, which is operated by the Association of Universities for Research in Astronomy, Inc., for NASA, under contract NAS5-26555. O.G. is supported by the European Research Council under Grant No. 742104 and by the Swedish Research Council (VR) under the grants 2018-03641 and 2019-02337. T.L. is supported by the Swedish Research Council under contracts 2019-05135 and 2022-04283 and the Swedish National Space Agency under contract 117/19. C.B. and T.L. are also supported by the European Research Council under grant 742104.
S.J. acknowledge support by the Vetenskapsr{\aa}det (Swedish Research Council) through contract No.  638-2013-8993 and the Oskar Klein Centre for Cosmoparticle Physics.

\bibliography{main}

\end{document}